\documentclass[a4paper]{jpconf}

\usepackage[title,titletoc]{appendix}
\usepackage[utf8]{inputenc}
\usepackage[square,sort,comma,numbers]{natbib}
\usepackage{graphicx}
\usepackage{lineno}
\usepackage[breaklinks=true]{hyperref}
\usepackage{hypernat}
\usepackage{epsfig}
\usepackage{wrapfig}

\usepackage{graphicx,type1cm,eso-pic,color}

\usepackage{fancyhdr}
\usepackage{pslatex}   
\usepackage{color}
\usepackage{amsmath, amsthm, amssymb} 
\usepackage{endnotes}
\usepackage{xspace}
\usepackage{multirow}

\usepackage{libertine}

\begin{document}
\title{Studies of final state interactions via femtoscopy in ALICE}

\author{\L{}ukasz Kamil Graczykowski (for the ALICE Collaboration)}

\address{Faculty of Physics, Warsaw University of Technology, ul. Koszykowa 75, 00-662, Warszawa, Poland}

\ead{lgraczyk@if.pw.edu.pl}

\begin{abstract}
Femtoscopy is a technique enabling measurements of the space-time characteristics of particle-emitting sources. 
However, the femtoscopic analysis is also sensitive to the interaction cross-section. In this paper we show the first preliminary measurements of $\rm K^0_SK^{\pm}$ correlation functions in Pb--Pb collisions at $\sqrt{s_{\rm NN}}=5.02$~TeV. These correlations originate from the final-state interactions which proceed through the $a_0(980)$ resonance only and can be employed to constrain its parameters. A similar approach can be applied to baryon pairs to extract the unknown interaction cross-sections for some \mbox{(anti-)baryon--(anti-)baryon} pairs. We show baryon--baryon and baryon--anti-baryon correlation functions of protons and lambdas, as well as discuss shortly the fitting method.

\end{abstract}

\section{Introduction}
The ``traditional'' method of femtoscopy (also known as ``HBT" correlations), created to measure the space-time characteristics of the particle-emitting region at freeze-out, is based on measurements of correlations of two identical bosons (usually pions). These correlations are calculated as a function of relative momentum, expressed as $q=p_1-p_2$ or $k^{\ast}=q/2$~\cite{Goldhaber:1960sf,Lisa:2005dd}. In general, the femtoscopic technique allows the width of the distribution of relative separation between the emission points of two particles, usually called the ``HBT radius", to be extracted. The dependence of this parameter on event multiplicity (centrality) and transverse momentum of the pair $k_{\rm T}=|p_1+p_2|/2$ is frequently interpreted in terms of relativistic hydrodynamics as a signature of collective behavior of the created medium~\cite{Lisa:2005dd}. The mathematical formalism of femtoscopy is not limited only to identical bosons though and other pair combinations can be studied as well. The results of the femtoscopic analysis of two such not typical pair combinations are presented below. 

\section{Femtoscopy of $\mathrm{K}^0_{\rm S}\mathrm{K}^{\pm}$ pairs}

The results of charged and neutral femtoscopy of identical kaons have been published by ALICE in Pb--Pb and pp collisions~\cite{Abelev:2012ms,Abelev:2012sq,Adam:2015vja}. In the case of identical kaons, the correlation effect at low relative momentum originates from the combination of the following physical effects: Bose-Einstein quantum statistics (for $\mathrm{K}^{\pm}\mathrm{K}^{\pm}$ and $\mathrm{K}^0_{\rm S}\mathrm{K}^0_{\rm S}$), Coulomb interaction (for $\mathrm{K}^{\pm}\mathrm{K}^{\pm}$), and final-state interaction (FSI) via the $f_0(980)/a_0(980)$ threshold resonances (for $\mathrm{K}^0_{\rm S}\mathrm{K}^0_{\rm S}$).

The non-identical correlations of $\mathrm{K}^0_{\rm S}\mathrm{K}^{\pm}$ have never been studied before. In this case the FSI goes through one resonant state of $a_0(980)$\footnote{Both the $\mathrm{K}^0_{\rm S}\mathrm{K}^0_{\rm S}$ pair and the $a_0$ are in $I=1$ isospin state. On the other hand, the $f_0$ is in $I=0$ state, so this channel is not allowed since the isospin would not be conserved.}. This uniqueness of the $\mathrm{K}^0_{\rm S}\mathrm{K}^{\pm}$ pair allows the properties of the $a_0$ resonance itself to be constrained, which is discussed in many publications as a tetraquark (a 4-quark state), or a ``$\rm \overline{K}$-$\rm K$ molecule". In this study we considered four $a_0$ parameterizations: ``Achasov2"~\cite{Achasov:2002ir}, ``Achasov1"~\cite{Achasov:2001cj}, ``Antonelli"~\cite{Antonelli:2002ip}, and ``Martin"\cite{Martin:1976vx}.



\begin{figure}[!hbt]
	\centering
	\includegraphics[width=0.6\textwidth]{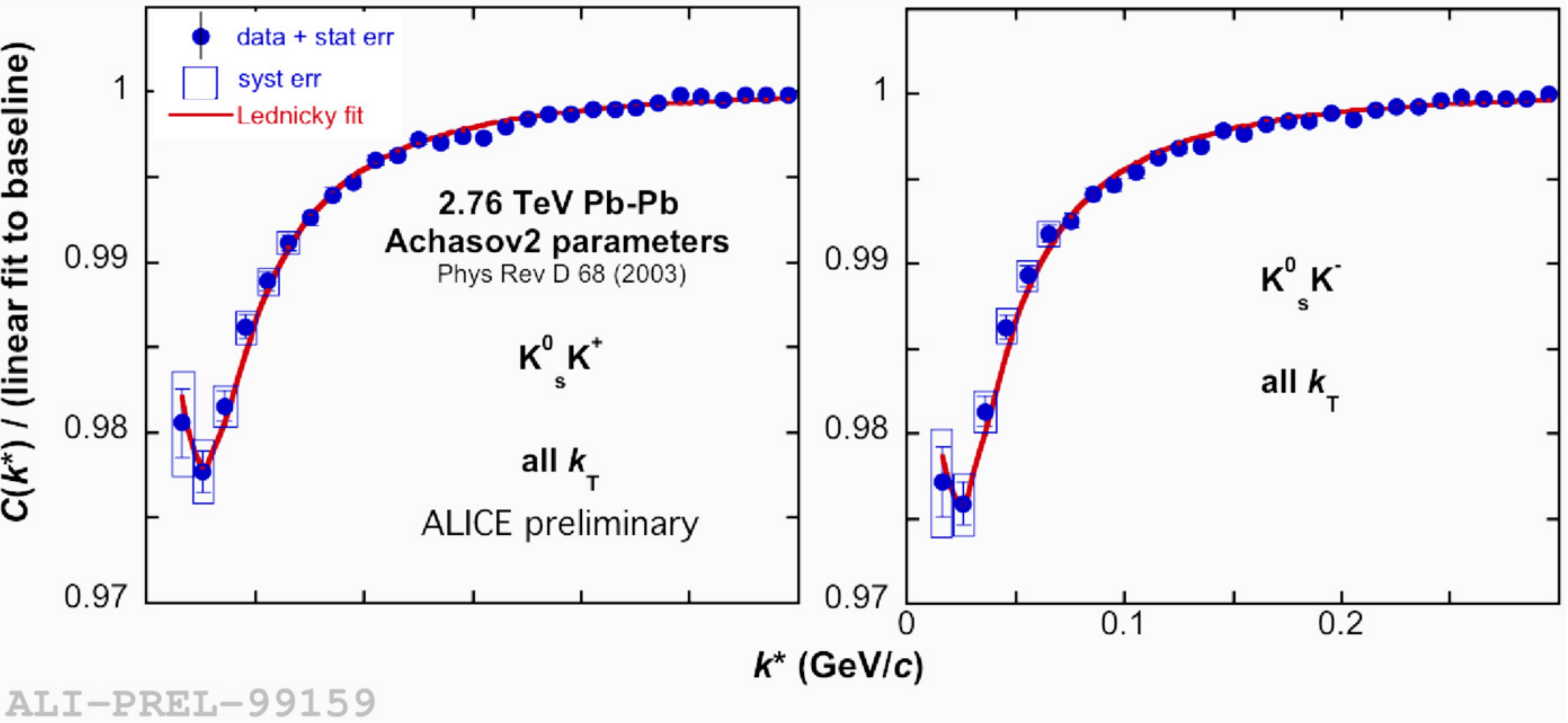}
	\caption{\label{fig:K0sK+-}
	$\mathrm{K^0_S}\mathrm{K^+}$ (left) and $\mathrm{K^0_S}\mathrm{K^-}$ (right) correlation functions with Lednicky fits using "Achasov2" parameters of $a_0$.
	}
\end{figure}

Examples of $\mathrm{K^0_S}\mathrm{K^+}$ and $\mathrm{K^0_S}\mathrm{K^-}$ correlation functions measured by ALICE in two $k_{\rm T}$ ranges, plotted together with fits of the so-called Lednicky analytical model~\cite{Lednicky:1981su,Bekele:2007zza} emplying the ``Achasov2" parameters of $a_0$, can be seen in Fig.~\ref{fig:K0sK+-}. One can observe a suppression at low $k^{\ast}$ which is a result of the strong FSI. By looking at the plots themselves one can judge that the fit with $a_0$ FSI gives a very good description of the experimental data.

\begin{figure}[!hbt]
	\centering
	\includegraphics[width=0.495\textwidth]{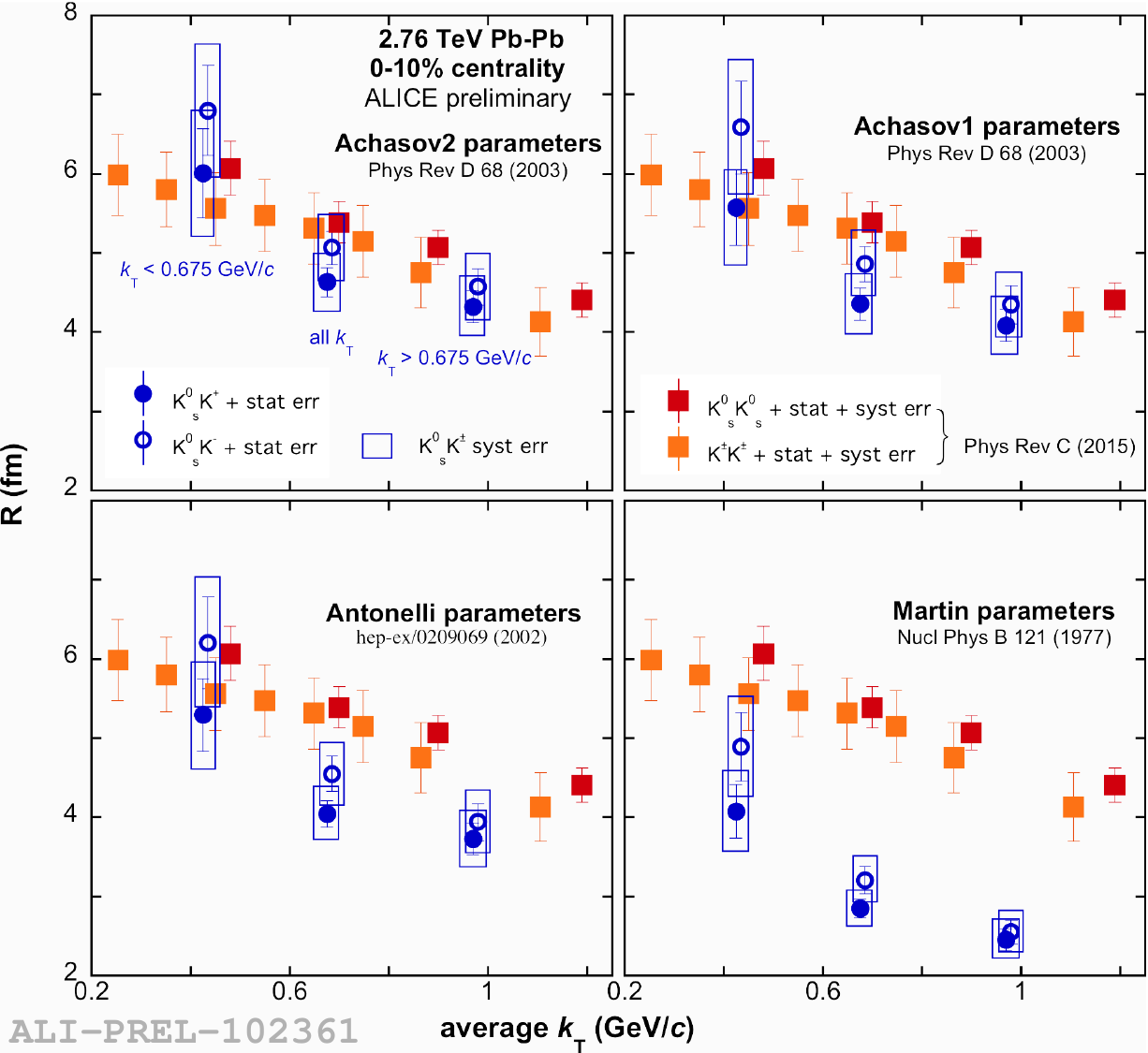}
	\hfill 
	\includegraphics[width=0.495\textwidth]{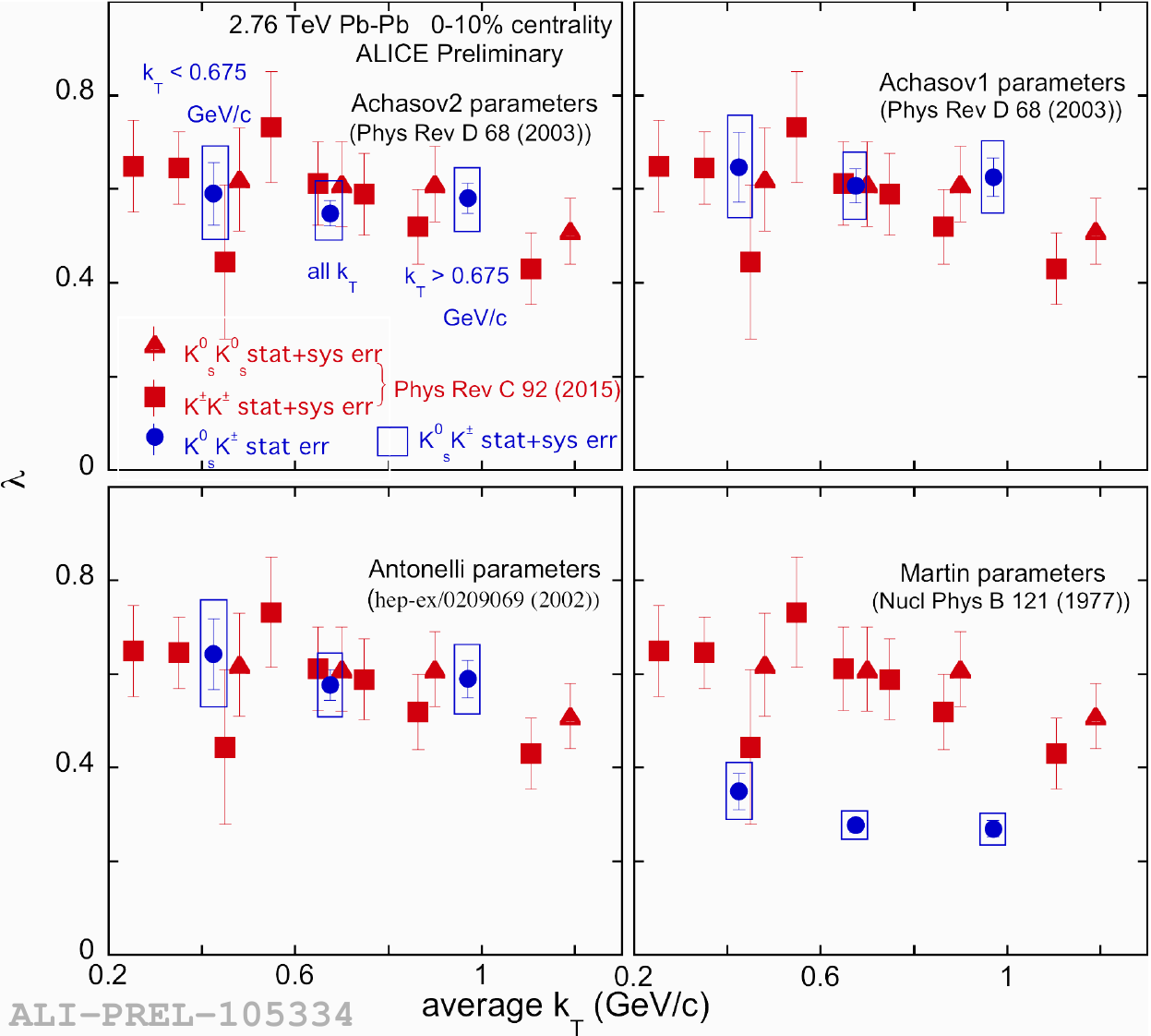}
	\caption{\label{fig:RlamK0sK+-}
	$R$ (left) and $\lambda$ (right) parameters of the fit from averaged $\mathrm{K}^{0}_{\rm S}\mathrm{K}^{\pm}$ correlations compared to the results of identical-kaon femtoscopy from ALICE~\cite{Adam:2015vja}.
	}
\end{figure}

The ``Lednicky'' fits allowed typical femtoscopic quantities, that is the ``HBT radius'' (the kaon source size) $R$ and the correlation strength $\lambda$ to be extracted for all $a_0$ parameters. It was found that for all parameterizations the fit results for $\mathrm{K^0_S}\mathrm{K}^{+}$ and $\mathrm{K^0_S}\mathrm{K}^{-}$ are consistent with each other, therefore both quantities were averaged. Both $R$ and $\lambda$ are presented in Fig.~\ref{fig:RlamK0sK+-}.

The results show a clear agreement between identical-kaon and $\mathrm{K^0_S}\mathrm{K}^{\pm}$ results for all $a_0$ parameters except ``Martin'' which are the lowest ones in comparison to other considered parameterizations. Since there is no physical reason to expect that femtoscopic quantities for $\mathrm{K^0_S}\mathrm{K}^{\pm}$ pairs should be different from those for $\mathrm{K}^{0}_{\rm S}\mathrm{K}^{0}_{\rm S}$ and $\mathrm{K}^{\pm}\mathrm{K}^{\pm}$ pairs, the experimental data seems to favor higher values of $a_0$.

\section{Femtoscopy of baryon pairs}
Femtoscopy of baryons also allows to study final-state interactions. For example, FSI effects can play a role in the rescattering phase for final-state hadrons and, going beyond the relativistic heavy-ion field, in understanding the properties of neutron stars~\cite{SchaffnerBielich:2008kb,Wang:2010gr} or search for potential CPT symmetry breaking~\cite{Adamczyk:2015hza}. It is argued that baryon--anti-baryon annihilatation can be responsible for the $\rm p/\pi^+$ ratio which falls short of thermal model expectations at the LHC~\cite{Preghenella:2012eu}. Several theoretical calculations argue that rescattering in the final-state should be taken into account in the measurements of particle yields~\cite{Werner:2012xh,Karpenko:2012yf,Steinheimer:2012rd}.

\begin{figure}[!hbt]
	\centering
	\includegraphics[width=0.32\textwidth]{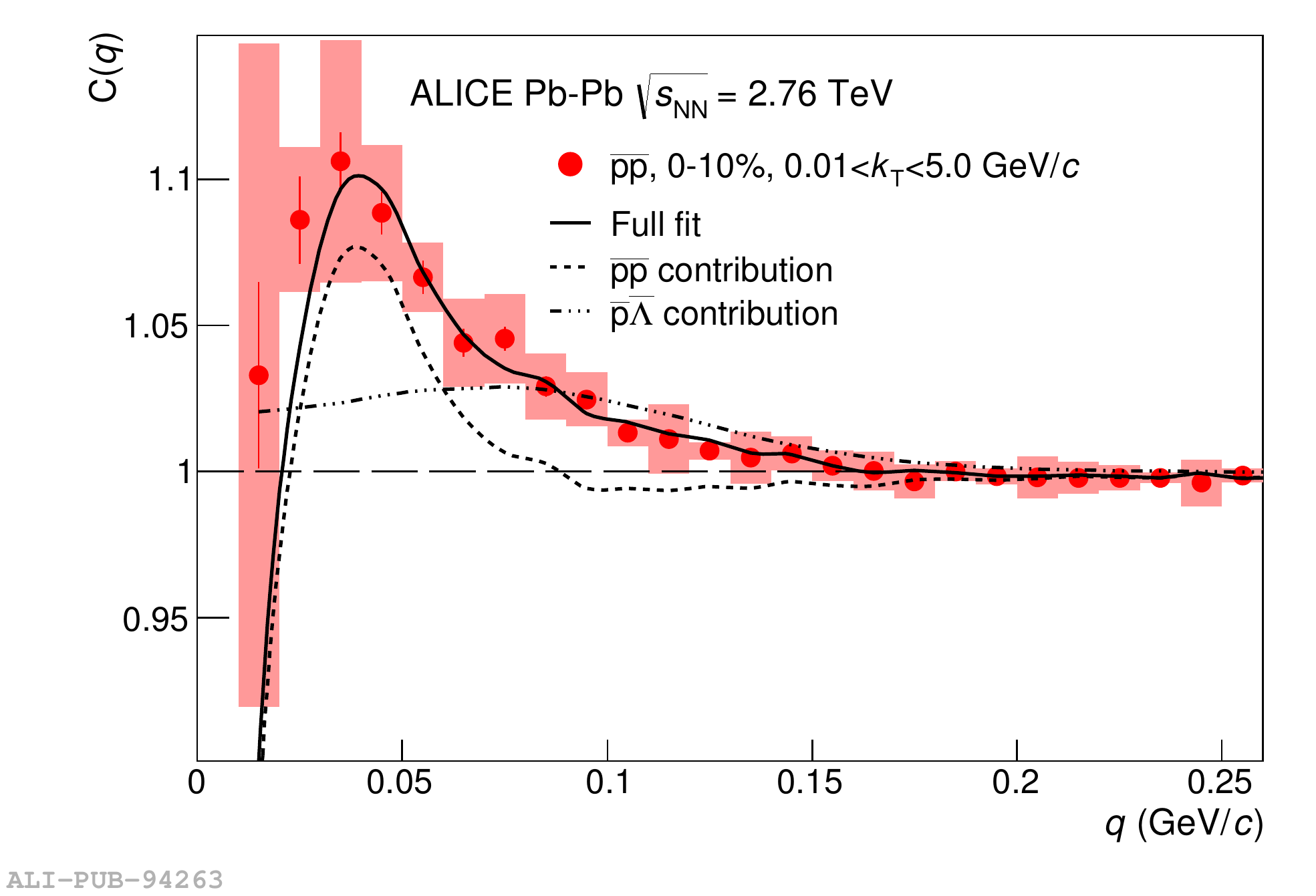}
	\hfill 
	\includegraphics[width=0.32\textwidth,height=3.7cm]{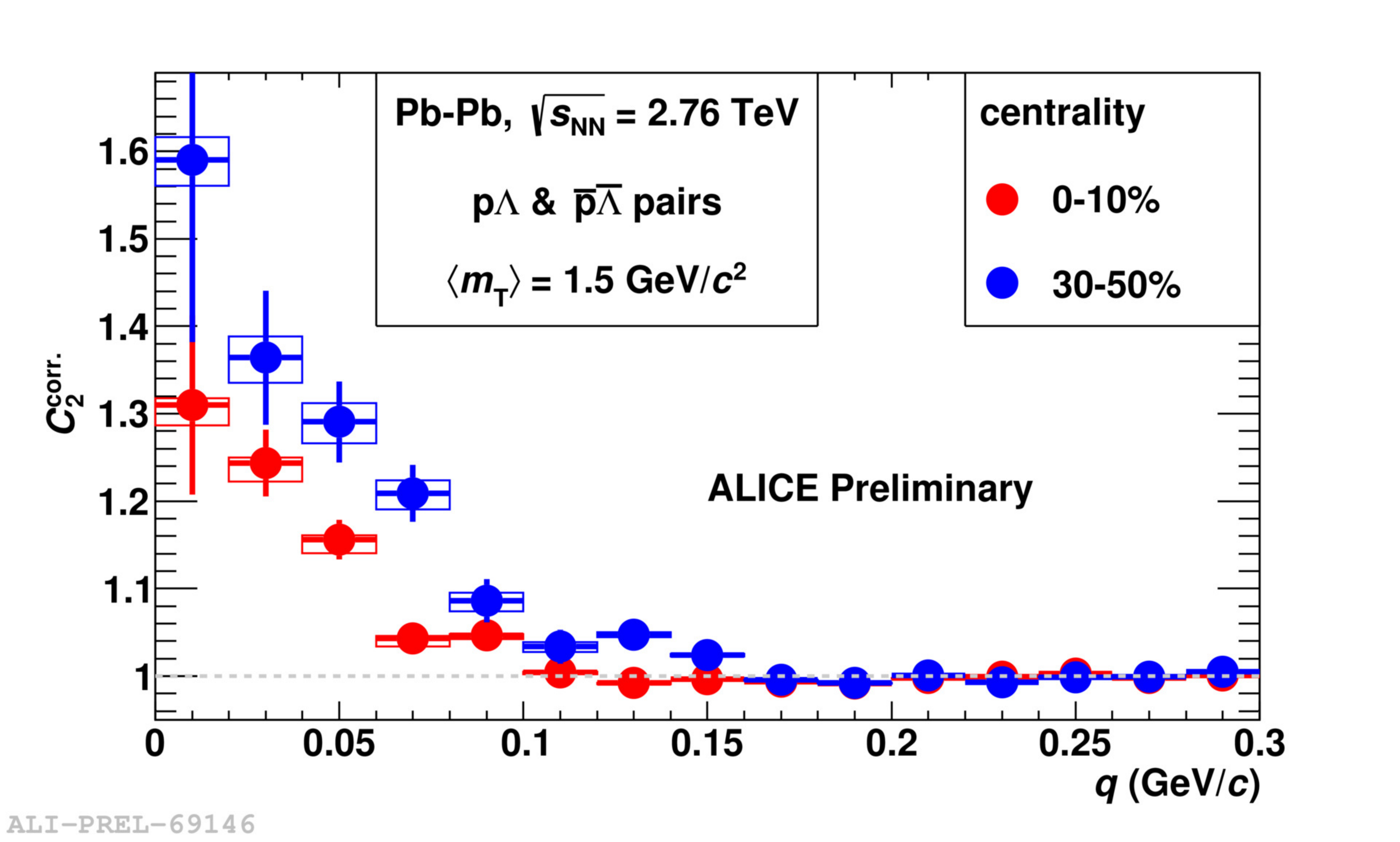}
	\hfill
	\includegraphics[width=0.32\textwidth]{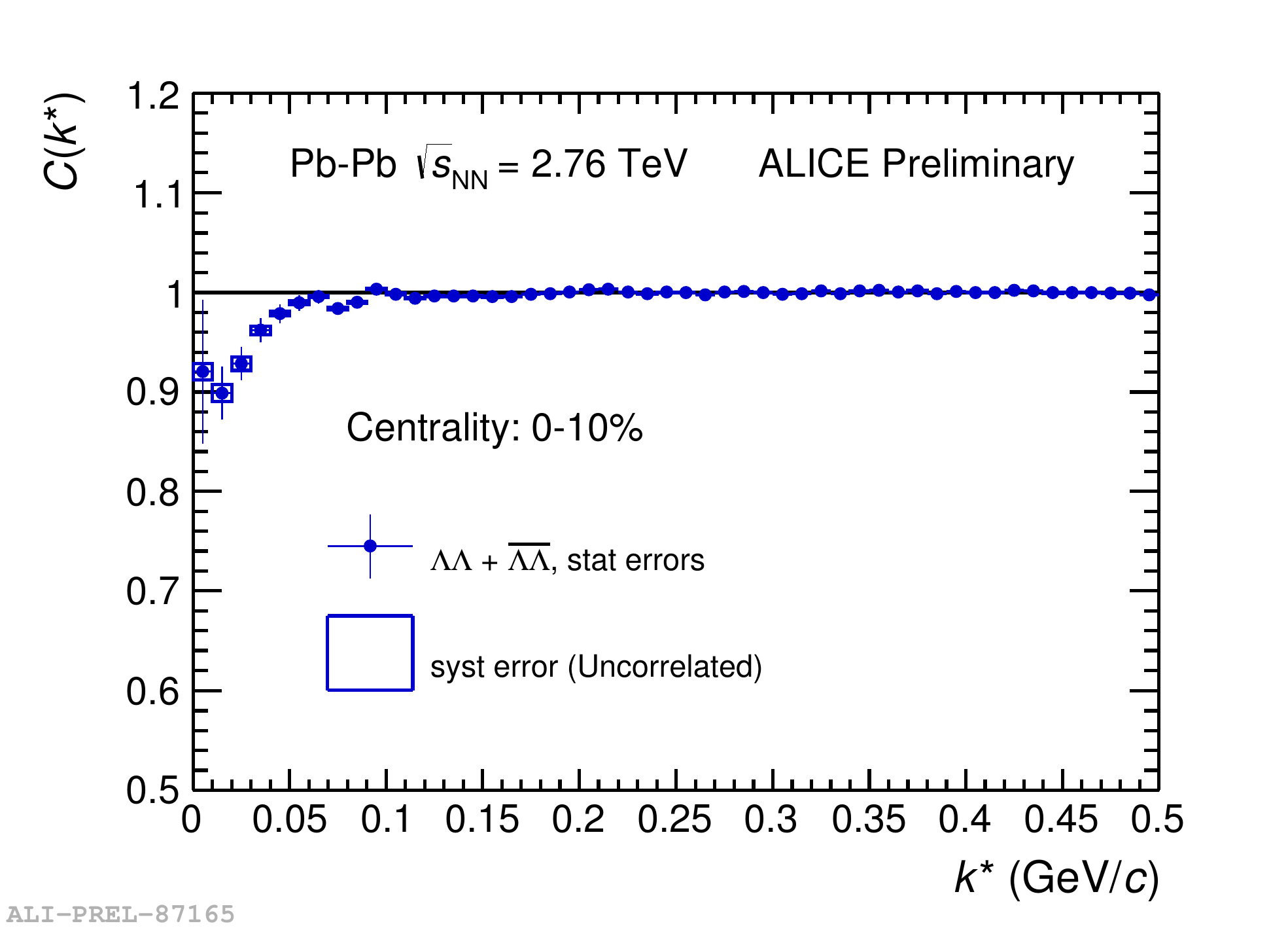}
	\includegraphics[width=0.32\textwidth]{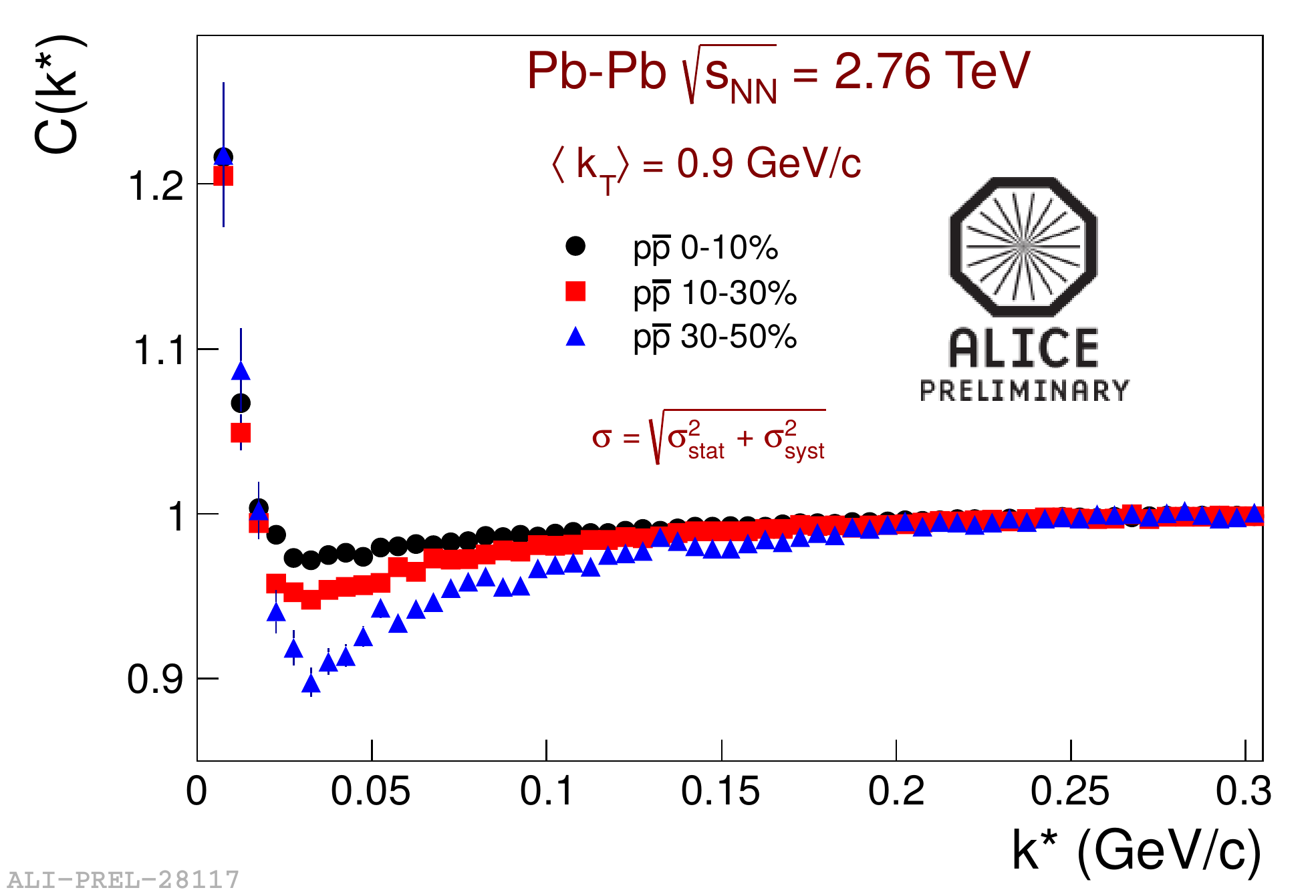}
	\hfill 
	\includegraphics[width=0.32\textwidth]{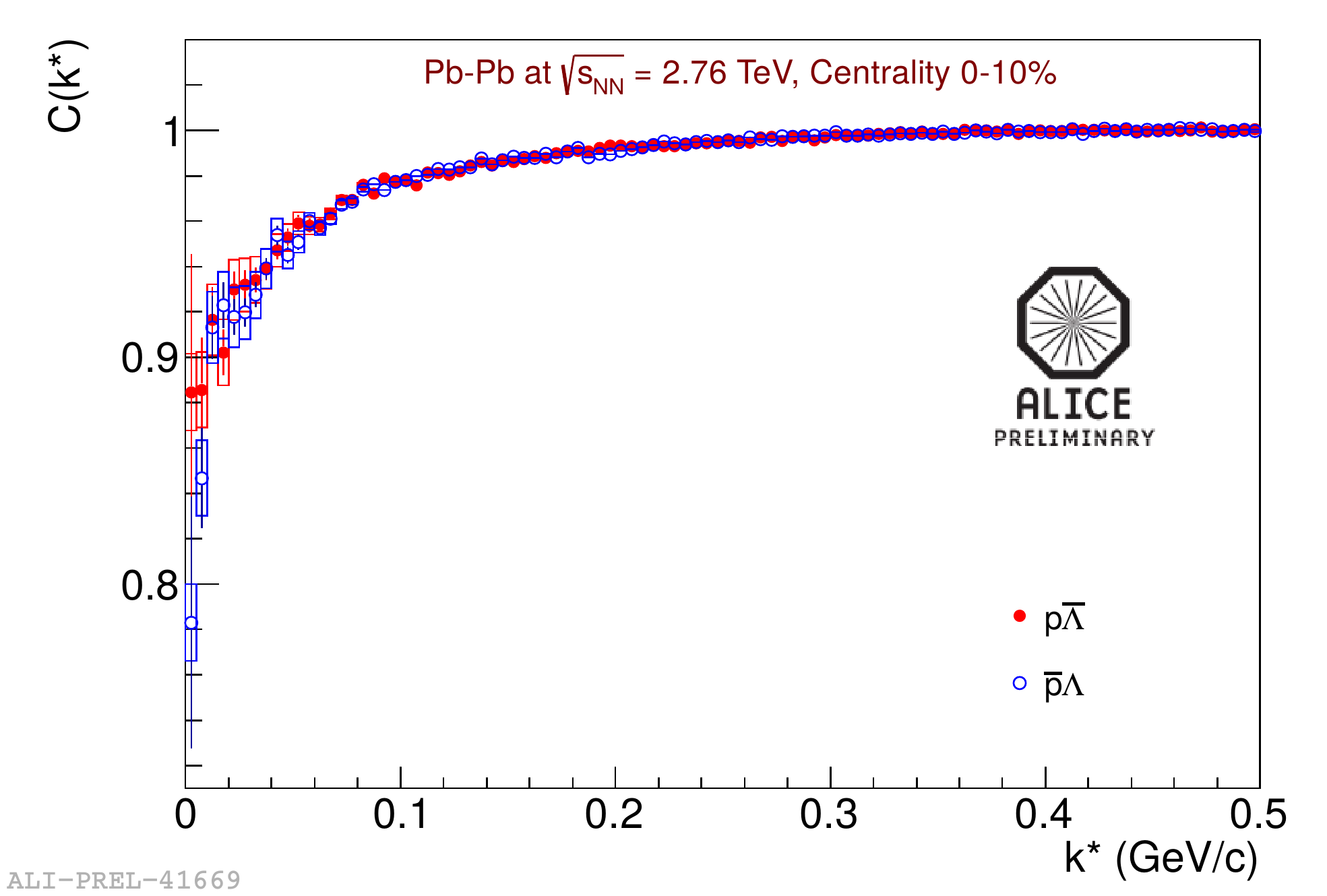}
	\hfill
	\includegraphics[width=0.32\textwidth]{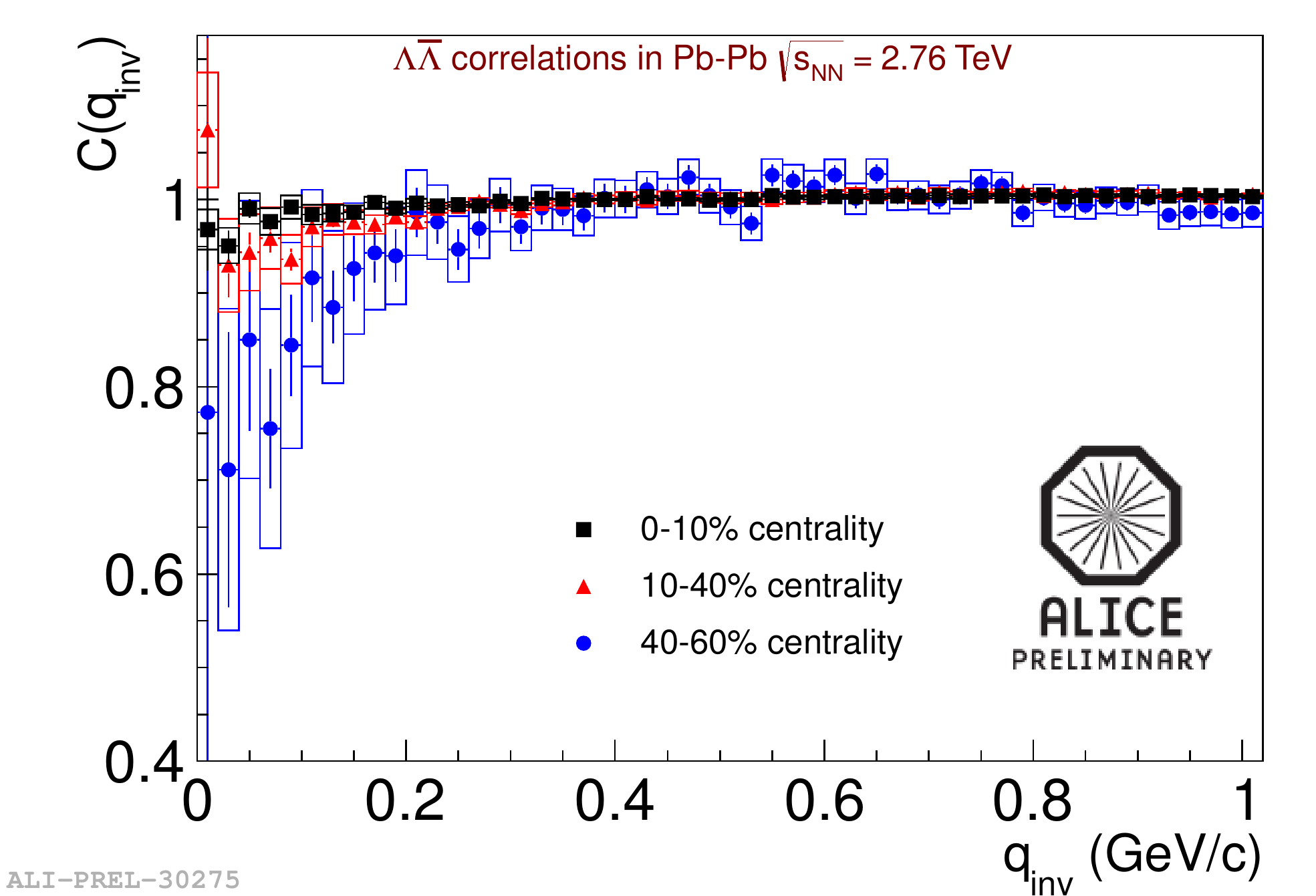}
	\caption{\label{fig:BBcorrs}
	Upper panel: correlation functions for (anti-)baryon--(anti-)baryon pairs: $\rm \overline{p}\overline{p}$ (left)~\cite{Adam:2015vja}, $\rm p\Lambda+\overline{p}\overline{\Lambda}$ (middle), and $\Lambda\Lambda+\overline{\Lambda}\overline{\Lambda}$ (right). Lower panel: correlation functions for baryon--anti-baryon pairs: $\rm p\overline{p}$ (left), $\rm p\overline{\Lambda}+\overline{p}\Lambda$ (middle), and $\rm p\overline{\Lambda}+\overline{p}\Lambda$ (right).
	}
\end{figure}

The ALICE experiment has already published results of $\rm pp$ and $\rm \overline{p}\overline{p}$ femtoscopy in Pb--Pb collisions at $\sqrt{s_{\rm NN}}=5.02$~TeV~\cite{Adam:2015vja}, which are complemented by still ongoing studies of other pairs with \mbox{(anti-)protons} and (anti-)lambdas. The upper panel of Fig.~\ref{fig:BBcorrs} shows baryon--baryon correlation functions for $\rm \overline{p}\overline{p}$, $\rm p\overline{\Lambda}+\overline{p}\Lambda$, $\Lambda\Lambda+\overline{\Lambda}\overline{\Lambda}$ and $\Lambda\overline{\Lambda}$ pairs. A clear correlation signal is present for all systems.  In the case of baryon--baryon correlations, the correlation effect at low relative momentum originates from the combination of the following phenomena: Fermi-Dirac quantum statistics (for $\rm pp+\overline{p}\overline{p}$ and $\Lambda\Lambda+\overline{\Lambda}\overline{\Lambda}$), Coulomb interaction (for $\rm pp+\overline{p}\overline{p}$), strong final-state interaction (for $\rm pp+\overline{p}\overline{p}$, $\Lambda\Lambda+\overline{\Lambda}\overline{\Lambda}$, and $\rm p\Lambda+\overline{p}\overline{\Lambda}$).

The $\rm \overline{p}\overline{p}$ correlation function exhibits a maximum around $q=40$~MeV/$c$ which is caused by the strong interaction, since both Coulomb and Fermi-Dirac statistics produce a negative correlation. Similar behavior, originating from these three effects is visible also for other pairs. In the case of $\rm p\Lambda+\overline{p}\overline{\Lambda}$, where only strong FSI is present, we do not observe any negative correlation. For $\Lambda\Lambda+\overline{\Lambda}\overline{\Lambda}$ the strong interaction gives surprisingly weak contribution since the correlation function at low relative momentum is clearly below unity due to Fermi-Dirac quantum statistics (though a hint of enhancement from strong FSI is visible for for the first bin).

In the lower panel of Fig.~\ref{fig:BBcorrs} correlation functions for baryon--anti-baryon pairs are presented. In this case only two effects can be present: Coulomb interaction (for $\rm p\overline{p}$) and strong FSI (for $\rm p\overline{p}$, $\rm p\overline{\Lambda}+\overline{p}\Lambda$, and $\rm p\overline{\Lambda}+\overline{p}\Lambda$). The Coulomb interactions is present only in the case of $\rm p\overline{p}$ pairs and, due to its attractive nature in case of opposite charges, produces a positive correlation. This effect is visible for the lowest values of $k^{\ast}$. Apart from this effect, all three pair combinations exhibit a wide suppression from strong FSI.

Fitting the baryon correlation functions is complicated by the presence of so-called ``residual correlations''. They arise from feed-down of weak decay products, which contaminate the experimental sample (e.g. an admixture of protons from decays of lambdas in the primary proton sample); see the $\rm \overline{p}\overline{p}$ correlation function, top-left plot in Fig.~\ref{fig:BBcorrs}. The fit with the $\rm \overline{p}\overline{p}$ wave function only cannot describe the experimental data. The visible excess is however accounted for if a contribution from $\rm \overline{p}\overline{\Lambda}$ pairs is considered. There are two methods incorporating ``residual correlations'' in the fitting procedure, ``transformed residuals''~\cite{Kisiel:2014mma} (used by ALICE in~\cite{Adam:2015vja}) and ``Gaussian residuals''~\cite{Shapoval:2014yha}.

\section{Conclusions}
Femtoscopy is a powerful tool capable not only to measure sizes of particle emitting sources but also gives access to final-state interactions. This feature allows to constrain strong interaction cross-sections for particle pairs where little or no other experimental data exist. The preliminary results from ALICE suggest that such detailed analysis for many meson and baryon pairs which has not been measured are possible at the LHC.
\\

The author wishes to acknowledge the financial support of the Polish National Science Centre under decisions Nos. DEC-2013/08/M/ST2/00598, DEC-2014/13/B/ST2/04054, and DEC-2015/19/D/ST2/01600.

\bibliographystyle{iopart-num}
\bibliography{bibliography}

\end{document}